\documentclass{article}

\usepackage{PRIMEarxiv}

\usepackage[utf8]{inputenc} 
\usepackage[T1]{fontenc}    
\usepackage{hyperref}       
\usepackage{url}            
\usepackage{booktabs}       
\usepackage{amsfonts}       
\usepackage{nicefrac}       
\usepackage{microtype}      
\usepackage{lipsum}
\usepackage{fancyhdr}       
\usepackage{graphicx}       
\graphicspath{{media/}}     
\usepackage{amsmath}

\title{Building Semantic Communication System via Molecules: An End-to-End Training Approach
}

\author{
  Yukun Cheng, Wei Chen, Bo Ai\\
  School of Electronic and Information Engineering \\
  Beijing Jiaotong University \\
  Beijing, China \\
  \texttt{\{ykcheng, weich, boai\}@bjtu.edu.cn} \\
}

\begin{document}
\maketitle

\begin{abstract}
The concept of semantic communication provides a novel approach for applications in scenarios with limited communication resources. In this paper, we propose an end-to-end (E2E) semantic molecular communication system, aiming to enhance the efficiency of molecular communication systems by reducing the transmitted information. Specifically, following the joint source channel coding paradigm, the network is designed to encode the task-relevant information into the concentration of the information molecules, which is robust to the degradation of the molecular communication channel. Furthermore, we propose a channel network to enable the E2E learning over the non-differentiable molecular channel. Experimental results demonstrate the superior performance of the semantic molecular communication system over the conventional methods in classification tasks.
\end{abstract}

\keywords{Semantic communication \and End-to-end learning \and Molecular communication \and Deep learning}

\section{Introduction}
Under the guidance of Shannon's Information Theory \cite{1}, nearly all current communication systems, including fifth-generation communication systems, aim to reliably transmit bit packets to corresponding receivers within specific constraints. The optimization objectives in these systems mainly focus on efficiently transmitting bits without considering the specific usages of these bits in upper-layer applications. However, emerging applications in the next-generation mobile communication systems \cite{2}, such as industrial robots, digital twins, smart cities, etc., exhibit different requirements and constraints. In these scenarios, received signals are utilized for inference tasks instead of reconstructing the original data, thereby necessitating a transformation in the content of transmitted data. 
Semantic communications \cite{3, 10480348, 10107616}, where task-relevant information is taken into account during data processing, align with the requirements mentioned above. Under this new communication paradigm that considers semantics and pragmatics, we can significantly reduce the traffic load, and provide potential solutions for scenarios with strict constraints on communication resources.

Molecular communication is the most widely employed communication mechanism on Earth \cite{4}. For example, insects use information molecules known as pheromones to transmit and receive information, coordinating group activities. Inspired by nature, researchers utilize chemical molecules, called information molecules, for information transmission through the release, propagation, and reception of these molecules \cite{5, 6}. This synthetic molecular communication system was proposed as an alternative communication paradigm, intended to address scenarios where electromagnetic waves (EM)-based communication is impractical or inconvenient. For instance, the use of EM-based communication systems situated in confined environments, such as underground, tunnels, pipelines, or saltwater environments, can be highly inefficient \cite{7,8,9}, where electromagnetic waves exhibit higher propagation loss over long distances. Furthermore, when dealing with situations that involve limited dimensions, such as communication within the human body and the operation of nano-scale robots \cite{10,11,12}, the utilization of devices such as antennas is restricted due to size constraints. Nevertheless, a majority of existing molecular communication systems encounter limitations in terms of low bandwidth and communication rates, and high inter-symbol interference (ISI) \cite{5}. Previous studies have primarily concentrated on developing novel modulation techniques \cite{13} and channel coding methods \cite{14} to enhance the performance of communication links. However, there has been limited attention given to optimizing the entire system from the perspective of the information source. The concept of semantic communication shows the potential to significantly reduce the amount of information that needs to be transmitted, thereby improving the efficiency of molecular communication.

The design principles of semantic communication are partly in line with joint source-channel coding (JSCC) \cite{9438648, 10225310, 9954153, 15, 16}, where source signals are directly mapped to channel symbols and estimates are recovered at the receiver. The application of deep neural networks in the design of communication systems has become widespread owing to their ability to effectively fit data and perform nonlinear transformations. Introduced by \cite{15}, DeepJSCC employs deep neural networks as encoders and decoders, forming an autoencoder optimized jointly in a data-driven manner. The encoders and decoders are trained to extract meaningful semantic features for end-to-end communication. In this type of communication, the information learned is transmitted based on the underlying task \cite{16}. Despite the success of deep learning-based JSCC (DJSCC) in achieving effective semantic communication, current frameworks are primarily designed for differentiable channel scenarios, such as traditional EM-based communication with the additive white Gaussian noise (AWGN) or fading channels which allow gradient backpropagation for joint optimization. However, in a molecular communication system, where information propagates through information molecules, gradient backpropagation is typically not feasible.

The problem of enabling end-to-end (E2E) learning in model-free wireless channels was first investigated in \cite{20}. There are currently two main approaches. One involves learning a differentiable channel model using a neural network, such as Generative Adversarial Networks (GANs) \cite{17} or their variants \cite{18}, and Gaussian mixture networks \cite{19}. The channel network generates the channel output or simulates its conditional probability distribution, supporting gradient backpropagation while providing symbols for the receiver. The other approach aims to bypass the non-differentiable channel through gradient approximation \cite{20}. The receiver is trained with the true gradient, while the transmitter is trained using a gradient approximation obtained by sampling the channel. While \cite{18} offers a training approach for molecular communication, the GANs-based method cannot provide an explicit channel model, which is important for the networks to learn a robust signal representation under high dynamic channel conditions.

In this paper, we present a JSCC approach for a semantic molecular communication system. We treat the encoding, propagation, and inference of the source as an E2E semantic communication task, where the molecular communication channel model is considered in the transmission, propagation, and reception of information molecules. To the best of our knowledge, this is the first work that introduces semantic communication in molecular communication systems. Furthermore, we introduce a novel channel network by incorporating the unique characteristics of molecular communication, which enables E2E learning in the non-differentiable molecular propagation channels. Overall, the contributions of this paper are as follows:
\begin{itemize}
    \item We present a novel end-to-end (E2E) semantic molecular communication system, which to the best of our knowledge, is the first AI-based E2E molecular communication system that considers tasks involving source information. 
    \item We propose a channel network to support the E2E training under the non-differentiable molecular channel. 
\end{itemize}

\section{Preliminary}\label{s2}
\subsection{System Model}\label{s21}
\begin{figure}
    \centering
    \includegraphics[width=0.7\textwidth]{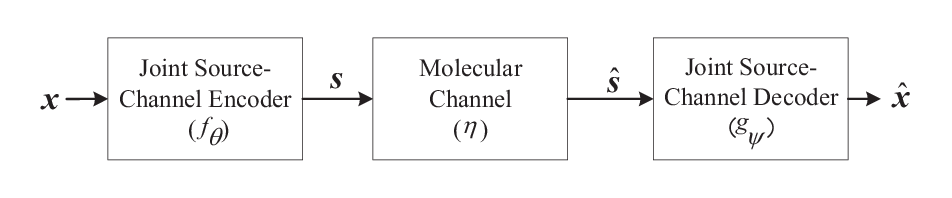}
    \caption{The model of the point-to-point semantic molecular communication system.}
    \label{System Model}
\end{figure}
In this work, we consider a point-to-point molecular semantic communication system as shown in Figure~\ref{System Model}, and typical image inference tasks in the experiments. An input image of size $H(height) \times W(width) \times C(channel)$ is represented by a vector $\boldsymbol{x}\in\mathbb{R}^n$, where $n =H\times W \times C$ and $\mathbb{R}$ denotes the set of real numbers. The joint source-channel encoder encodes $\boldsymbol{x}$ and the encoding function $f_\theta(\boldsymbol{x})$ returns a vector of channel input symbols $\boldsymbol{s} \in \mathbb{R}^k$. The encoding process can be expressed as:
\begin{equation}
    \boldsymbol{s} = f_\theta(\boldsymbol{x}),
    \label{Encoding_euqation}
\end{equation}
where $k$ is the size of channel input symbols, $\theta$ is the parameter set of the encoder. Each component of $\boldsymbol{s}$ denotes the number of molecules transmitted in the slot. The number of molecules transmitted during each slot is limited to a minimum of $0$ and a maximum of $n_m$, following the power constraint in EM-based communications. 

The vector of the encoded symbol $\boldsymbol{s}$ is transmitted over a molecular propagation channel represented by the function $\eta(\cdot)$, which will be introduced shortly. The vector of the channel output symbols $\hat{\boldsymbol{s}} \in \mathbb{R}^k$ received by the decoder is processed by a decoding function $g_{\psi}(\cdot)$ as follows:
\begin{equation}
    y = g_\psi(\hat{\boldsymbol{s}}) = g_\psi(\eta(f_\theta(\boldsymbol{x}))),
    \label{Decoding_equation}
\end{equation}
where $y$ is the output of the decoder, and $\psi$ is the parameter set of the decoder. The loss function $\mathcal{L}$ is determined by the underlying task.
For example, in an image classification task, the loss function can be expressed as the cross-entropy (CE) written as follows:
\begin{equation}
    \label{CE}
    \mathcal{L}_{CE} = -\sum_{i} z_i \log(y_i),
\end{equation}
where $z_i$ and $y_i$ represent the one-hot encoding of the true label and the output of the decoder, respectively.

\subsection{Molecular propagation channel}\label{s22}
In a traditional molecular communication system, the transmitter encodes the signal to the properties of the information molecules, such as concentration, type of molecules, emission time, etc. For the simplest case, in a binary-concentration-shift-keying (BCSK) modulation scheme, each bit is transmitted during a time slot $t_s$. The transmitter releases $n_m$ molecules at the start of $t_s$ when observing bit $``1"$ and releases none for bit $``0"$. The receiver decodes the signal according to the number of received molecules. 

\begin{figure}
    \centering
    \includegraphics[width=0.5\textwidth]{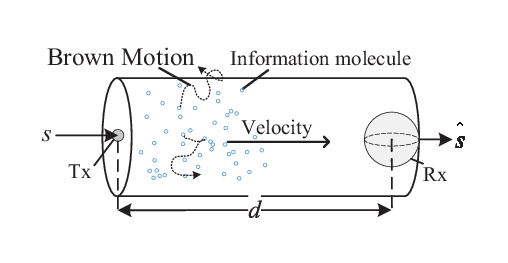}
    \caption{Molecular propagation channel.}
    \label{Molecular Model}
\end{figure}

In this work, we consider a single-input-single-output (SISO) molecular communication system with a pair of synchronized point transmitter and spherical passive receiver as shown in Figure~\ref{Molecular Model}, where the information is encoded into the concentration of the molecules. Assuming that the transmitter can release a certain number of information molecules at the same time, which share identical shape and size, and have no interactions and elastic collisions between each other.
All the information molecules propagate in a three-dimensional cylindrical pipe where the medium has a constant uniformly positive flow. The transmitter is located at a distance $d$ from the center of the spherical receiver, and the radius of the receiver is $r$.

The coordinates of the transmitter and receiver under 3D Cartesian coordinate system are set as follows: $\textbf{d}_{Tx} = [0,0,0]$ for the transmitter, $\textbf{d}_{Rx} = [d,0,0]$ for the center of the receiver, and $\textbf{d}_m(t) = [x_t, y_t, z_t]$ for the position of an information molecule at time $t$. The velocity vector of the flow is $\boldsymbol{v}=[v,0,0]$. The diffusion coefficient for the uncertain Brown motion of the molecules is $D_c$.
Let $\emph{c}(\textbf{d}, t)$ denote the concentration of molecules, i.e., the average number of molecules at coordinate $\textbf{d}$ and time $t$. The propagation process of molecules can be characterized by the advection-diffusion equation \cite{21}:
\begin{equation}
\label{advection-diffusion}
    \frac{\partial c (\textbf{d}, t)}{\partial t} = D_c \bigtriangledown^2 c(\textbf{d}, t) - \bigtriangledown (\boldsymbol{v} \cdot c(\textbf{d},t)), 
\end{equation}
where $\bigtriangledown^2$ and $\bigtriangledown$ are the Laplace operator and the Hamiltionian in Cartesian coordinates, respectively. 
Given Eq. \ref{advection-diffusion}, the probability of observing one molecule per unit volume at the coordinate $\textbf{d}$ and time $t$ can be calculated according to \cite{21}, which leads to
\begin{equation}
    \label{PDF}
    f_m(\textbf{d}, t) =  \frac{1}{{4\pi D_c t}^{3/2}}  exp\left(-\frac{||\textbf{d} - t \boldsymbol{v} - \textbf{d}_{Tx}||_2^2}{4 D_c t}\right).
\end{equation}

According to the Uniform Concentration Assumption \cite{21}, when the receiver is placed far away from the transmitter, the probability of observing a molecule for all points within the receiver volume can be approximated to the value at the center of the receiver. Thus, by inserting the position of the transmitter and receiver $\textbf{d}_{Tx}$ and $\textbf{d}_{Rx}$ to Eq. \ref{PDF}, the probability of observing one molecule by the transceiver can be written as
\begin{equation}
    P(t) = \frac{V_r}{(4\pi D_c t)^{3/2}}  exp\left(-\frac{(d-tv)^2}{4D_c t}\right),
\end{equation}
where $V_r$ is the volume of the spherical receiver and $v$ is the x-axis velocity of the flow.

For the conventional BCSK modulation scheme, the number of the molecules observed by the receiver, denoted by $R_m$, follows a binomial distribution. When $n_m$ is sufficiently large, the distribution becomes a Gaussian distribution.
The distribution of the number of molecules received by the receiver can be rewritten as
\begin{equation}
    \label{distribution}
    \begin{aligned}
        R_m(n_m,t) &\sim \mathbf{\mathcal{B}} (n_m, P(t)) \\ &\sim \mathcal{N}(n_mP(t), n_mP(t)(1-P(t))),
    \end{aligned}
\end{equation}
where $\mathcal{B}$ and $\mathcal{N}$ denotes the binomial distribution and the Gaussian distribution, respectively.

During the communication process, the molecules emitted in previous time slots are received by the receiver due to the uncertain diffusion of the molecules. The influence of this inter-symbol-interference (ISI) on molecular communication systems is negligible when the flow velocity significantly exceeds the Brownian motion. However, when diffusion becomes the predominant mode of propagation for molecules, this interference can lead to severe performance degradation in communication. Meanwhile, there may be some Gaussian channel noise originating from molecular decomposition or other nano-machine emissions, denoted as $N_{noise} \sim \mathcal{N}(0, \sigma_n^2)$.
For the transmitted bits $\boldsymbol{b}$, the number of molecules observed by the receiver at the $j$-th time slot can be expressed as
\begin{equation}
    N_{ob}(j,t) = \boldsymbol{b}[j]R_m(n_m,t) + \sum_{i=1}^l \boldsymbol{b}[j-i]R_m(n_m, t+it_s),
\end{equation}
where $l$ is the length of channel memory. For simplicity, we only consider the influence of the previous time slot in this paper, leading to $l=1$. 

To further quantify the effect of different system topological parameters and drift velocities on the number of arrived molecules, the signal-to-inference ratio (SIR) is defined as the ratio of received molecules released in the current time slot to the average number of molecules from the ISI and Gaussian noise, which is expressed as 
\begin{equation}\label{SIR}
    SIR = \frac{\boldsymbol{b}[j]R_m(n_m,t)}{(\boldsymbol{b}[j-1]R_m(n_m, t+t_s)+N_{noise})/2}.
\end{equation}
The increase in transportation distance $d$ and the decrease in flow velocity $v$ cause the degradation of SIR. A similar mathematical model is validated in \cite{18} through a particle-based simulation.

\section{The Proposed molecular semantic communication system}
\label{s3}
The existing research on the design of molecular communication systems mainly focuses on channel coding or modulation schemes, where the source signal and tasks are not taken into consideration. Here, we propose a JSCC approach for a semantic molecular communication system. We utilize deep neural networks for the extraction, encode/decode, and (de)modulation of the task-relevant features. Meanwhile, to enable E2E learning over the molecular propagation channel, we utilize a GM-based neural network to predict the distribution of the received signal and propagate gradients between the encoder and decoder.

\subsection{Network Architecture}

\begin{figure*}
    \centering
    \includegraphics[width=0.9\textwidth]{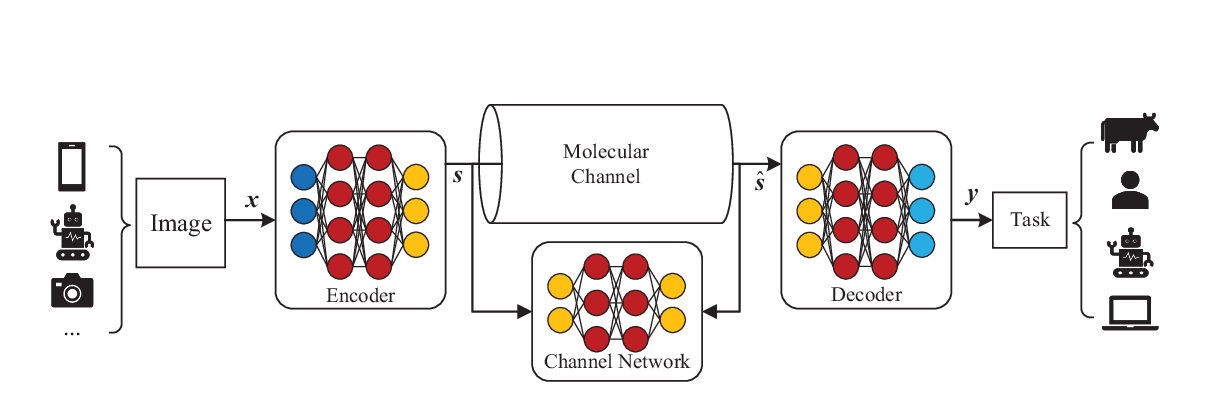}
    \caption{Architecture of the proposed system.}
    \label{Architecture}
\end{figure*}

The architecture of the considered communication system is shown in Fig. \ref{Architecture}. As described in Sec. {\ref{s21}}, an input image $\boldsymbol{x}$ is processed at the transmitter, where a deep neural network is employed as the source encoder, channel encoder, and modulator. 
The value of each element of the encoded symbol vector $\boldsymbol{s}$ represents the normalized number of molecules released in each time slot.
Information molecules are transmitted to the receiver through the molecular propagation channel described in Sec. \ref{s22}. At the receiver, the decoder calculates the number of detected information molecules at each slot, which leads to the received symbol vector $\hat{\boldsymbol{s}}$. Then, the joint source channel decoder utilizes $\hat{\boldsymbol{s}}$ to directly perform inference for the underlying task. Note that the proposed network could be applied to other modulation methods, via changing the output of the encoder to other molecular properties, e.g., emission time.

The goal of the E2E optimization is to find the optimal representation of task-specific information of the source under the unreliable molecular semantic channel. To achieve this, a deep neural network is used for the non-linear transformation at the encoder and decoder. Details of the network architecture of the encoder and decoder are provided in Table \ref{Network_architecture}, where the notation $\emph{F} \times \emph{F} \times \emph{K}$ denotes the convolutional layer has a kernel size of $\emph{F} \times \emph{F}$ and $\emph{K}$ filters. $c_{out}$ represents the number of filters of the last convolutional layer in the encoder.
The last layer of the encoder employs a Sigmoid activation to confine the value in the output symbol vector $\boldsymbol{s}$ to $[0,1]$, which is scaled by the maximum number of released molecules $n_m$ during the simulation of the molecular propagation channel.

\begin{table}[t]
    \centering
    \caption{Architecture of the Encoder and Decoder Layers}
    \label{Network_architecture}
        \centering
        \begin{tabular}{|c|c|c|c|c|c|}
            \hline
            EnLayer & Type & $\emph{F} \times \emph{F} \times \emph{K}$ & Stride & Padding & BatchNorm \\
            \hline
            Conv1 & Conv2d & 9x9x16 & 2 & 4 & BatchNorm2d \\
            Conv2 & Conv2d + LeakyReLU & 5x5x64 & 2 & 2 & BatchNorm2d \\
            Conv3 & Conv2d + LeakyReLU & 5x5x64 & 1 & 2 & BatchNorm2d \\
            Conv4 & Conv2d + LeakyReLU & 5x5x64 & 1 & 2 & BatchNorm2d \\
            Conv5 & Conv2d + Sigmoid & 5x5x$c_{out}$ & 1 & 2 & - \\
            \hline
        \end{tabular}
    \hspace{0.05\linewidth}
        \centering
        \begin{tabular}{|c|c|c|c|c|c|}
            \hline
            DeLayer & Type & $\emph{F} \times \emph{F} \times \emph{K}$ & Stride & Padding & BatchNorm \\
            \hline
            Conv1 & Conv2d + LeakyReLU & 5x5x64 & 1 & 2 & BatchNorm2d \\
            Conv2 & Conv2d + LeakyReLU & 5x5x64 & 1 & 2 & BatchNorm2d \\
            Conv3 & Conv2d + LeakyReLU & 5x5x64 & 1 & 2 & BatchNorm2d \\
            Conv4 & Conv2d + Sigmoid & 5x5x64 & 2 & 2 & - \\
            Linear & Linear & - & - & - & - \\
            \hline
        \end{tabular}
\end{table}

To train the network under the non-differentiable molecular propagation channel, we employ a channel network to model the conditional probability of the arrived molecules.
During the training stage, we use the network to generate the expected concentration of received molecules and propagate the gradient to the encoder.
During the inference stage, the molecules are emitted and detected by the point transmitter and spherical receiver through a molecule-based channel simulator. For demonstration purposes, readers can refer to \cite{citeone, citetwo, 9027995}, which is out of the scope of this manuscript.

\subsection{Channel Network}
Based on the probability of arriving information molecules, the number of received molecules follows a binomial distribution at the decoder as described in Sec. \ref{s22}. With a sufficiently large number of released molecules, the binomial distribution converges to a Gaussian distribution. As a result, the number of molecules that arrive at a specific moment can be expressed as a mixture of multiple Gaussian distributions. Each distribution represents the molecules released from different time slots before the current moment.

Inspired by \cite{19}, we propose a channel network to approximate the conditional distribution of the number of received molecules. This method is built based on mixture deep networks \cite{22}, which utilize deep neural networks to estimate the parameters of the conditional distribution of a point. The conditional probability density of the measured data is represented by
\begin{equation}
    p(\hat{\boldsymbol{s}}|\boldsymbol{s}) = \sum_{i=1}^h\pi_i(\boldsymbol{s})\phi_i(\hat{\boldsymbol{s}}|\boldsymbol{s}),
    \label{MDN}
\end{equation}
where $h$ is the number of mixture components, $\pi_i$ $(i=1,\cdots,h)$ are mixing coefficients of the probabilities associated with drawing a sample from each mixture component, and $\phi_i(\hat{\boldsymbol{s}}|\boldsymbol{s})$ is the kernel function representing the conditional density of $\hat{\boldsymbol{s}}$. $\pi_i$ satisfies $\sum_{i=1}^h \pi_i(\boldsymbol{s}) = 1$. Gaussian kernel functions are selected for estimating the detected molecules, which are given by
\begin{equation}
    \phi_i(\hat{\boldsymbol{s}}|\boldsymbol{s}) = \frac{1}{\sqrt{2\pi \sigma_i(\boldsymbol{s})^2}}exp \frac{-||\hat{\boldsymbol{s}}-\mu_i(\boldsymbol{s})||_2^2}{2 \sigma_i(\boldsymbol{s})^2},
    \label{kernel}    
\end{equation}
where $\mu_i$ and $\sigma_i^2$ are the mean and variance of each component in the mixture distribution, respectively.

\begin{figure}
    \centering
    \includegraphics[width=0.5\textwidth]{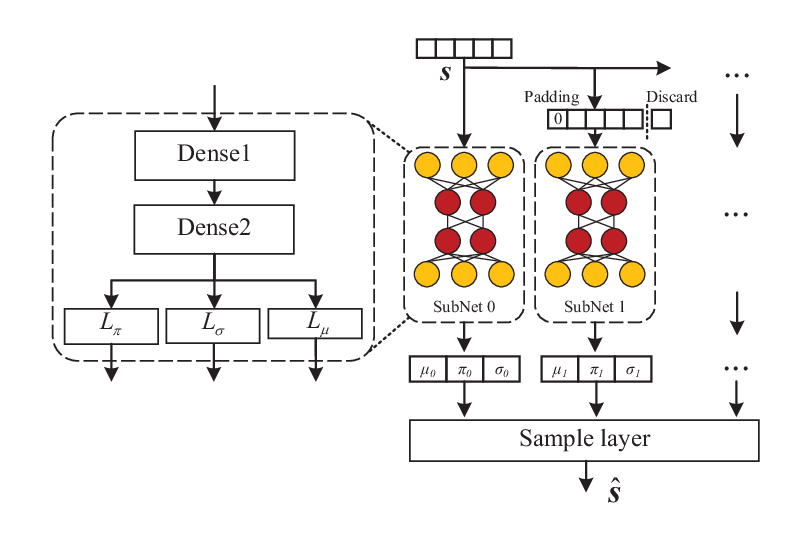}
    \caption{Architecture of the channel network}
    \label{Channel_Network}
\end{figure}
The architecture of our proposed channel network is given in Fig.~\ref{Channel_Network}. In a molecular communication system with uncertain Brown motion, the main cause of performance degradation is ISI, especially when the flow velocity is not significantly greater than the diffusion. Therefore, it is important to predict the arrival of molecules released in previous time slots. In the proposed network, we employ multiple subnetworks (SubNets) for predicting the number of arrived molecules released from previous time slots. For the $i$-th SubNet, the input vector is obtained by padding $i$ zeros to the start and discarding $i$ elements at the end of $\boldsymbol{s}$, accounting for molecules released from $i$-th slots earlier which bring ISI. The output of each SubNet is three vectors representing the mean, variance, and mixing coefficients of the distribution of the received molecule's number, respectively. All the evaluated parameters are sent into the sampling layer, which generates $\hat{\boldsymbol{s}}$ according to the combined distribution. For simplicity, we depict the architecture of a SubNet in Table \ref{Subnetwork}, where $h_{hidden}$ and $h$ are set as 20 and 2, respectively. The number of subnetworks is set to $l+1$, which is 2 in our study.

\begin{table}[t]
    \centering
    \caption{Parameters of Channel Subnets.}
    \label{Subnetwork}
    \begin{tabular}{|c|c|c|c|}
        \hline
        \textbf{Layer} & \textbf{Network Type} & \textbf{Input Dimension} & \textbf{Output Dimension} \\
        \hline
        Dense 1& Linear + LeakyReLU & 1 & $h\_hidden$ \\
        Dense 2& Linear + LeakyReLU & $h\_hidden$ & $h\_hidden$ \\
        $L_{\mu}$ & Linear & $h\_hidden$ & $h$ \\
        $L_{\sigma}$ & Linear + ReLU & $h\_hidden$ & $h$ \\
        $L_{\pi}$ & Linear + Softmax & $h\_hidden$ & $h$ \\
        \hline
    \end{tabular}
\end{table}

As shown in Fig.~\ref{Channel_Network}, in each SubNet, the channel symbol vector $\boldsymbol{s}$ is firstly processed by two full-connected linear layers with a leaky ReLU activation function, which is supposed to learn the features of the channel symbols. Then, the learned features are fed into three modules for parameter estimation, i.e., $`` L_\pi", ``L_\sigma"$ and $``L_\mu"$, which approximate the mixing coefficients $\pi_i$, variance $\sigma_i$, and mean $\mu_i$ of the distribution of the arrived molecules number released $i$ slots earlier, respectively. Each module has only one layer, whose structure is given in Table \ref{Subnetwork}. To satisfy the constraint of the mixing coefficients, i.e., $\sum_{i=1}^h \pi_i(\boldsymbol{s}) = 1$, a softmax function is applied at the end of the module $`` L_\pi "$. To avoid the negative value of the variance $\sigma$, a ReLU function is deployed for the module $`` L_\sigma "$.
During the training stage of the channel network, the parameters are utilized for computing the probability of predicting $\hat{\boldsymbol{s}}$, which is the loss function of the channel network, and will be detailed in Section \ref{s33}. In this process, sampling from the final distribution is not needed, which leads to the sample layer being inactivated. During the inference stage, the network outputs the sampled symbol vector from the evaluated channel distribution, given as $\hat{\boldsymbol{s}}$. 

The channel network is designed according to the characteristics of the molecular communication channels, structured with multiple sub-networks. The sub-networks estimate the impact of transmitted molecules in a time slot on the detected numbers of molecules in subsequent time slots. Meanwhile, the physical significance of the output of the network is the estimated mean and variance of the molecular concentration detected in each time slot.

\subsection{Training Method}\label{s33}
In this section, we will describe how to train the semantic molecular communication network with the assistance of the channel network.

The channel network and the semantic JSCC communication network are trained in order. We firstly train the channel network using randomly generated symbol vectors $\boldsymbol{s}$ and the corresponding received symbol vectors $\hat{\boldsymbol{s}} = \eta(\boldsymbol{s})$, which are generated using a simulator based on the molecular propagation channel described in Sec. \ref{s22}. 
The channel network is learned by maximizing the probability of predicting $\hat{\boldsymbol{s}}$ when injecting $\boldsymbol{s}$, which is equivalent to minimizing the loss function:
\begin{equation}
    \label{Channel_network_loss}
    \begin{aligned}
        \mathcal{L}_{CN} &= -\frac{1}{k} \sum_{j=1}^k log \left( p(\hat{\boldsymbol{s}}[j] | \boldsymbol{s}[j]) \right) \\
        &= -\frac{1}{k}\sum_{j=1}^k log\left(\sum_{i=1}^h \pi_i(\boldsymbol{s}[j]) \phi_i(\hat{\boldsymbol{s}}[j]|\boldsymbol{s}[j])\right),
    \end{aligned}
\end{equation}
where $\boldsymbol{s}[j]$ and $\hat{\boldsymbol{s}}[j]$ represent the transmitted and detected symbols at the $j$-th time slot, corresponding to the numbers of released and arrived information molecules, respectively. The loss function has been proven to be convex and to converge to the entropy of the target data distribution \cite{19}, i.e., the entropy of the probability distribution of $\hat{\boldsymbol{s}}$. After the training, the parameters of the channel network are frozen and will not be updated for the rest of the training stage.

Then we train the semantic JSCC communication network, where the encoder encodes the image $\boldsymbol{x}$ into channel symbols $\boldsymbol{s}$, and the decoder processes $\hat{\boldsymbol{s}}$ and outputs the semantic information $\boldsymbol{y}$ for some specific task. The input of the decoder, which is the channel network output $\hat{\boldsymbol{s}}$, is obtained by sampling the evaluated distribution. The encoder and decoder are trained jointly, with a fixed channel network. 

\section{Experiments}
\label{SIMULATION-RESULTS}

In this section, we verify the idea of the proposed channel network and evaluate the proposed semantic molecular communication system under two different scenarios.
The settings of the molecular propagation channel are listed in Table \ref{Molecular_channel_parameters}, where different distances and flow velocities are considered in the two scenarios. Scenario 1 considers short-distance molecular communications with a slow flow velocity, such as biological in-body communication applications \cite{24}. Scenario 2 considers long-distance communications with a fast flow velocity, such as industrial facilities, oil pipelines, and other applications in confined environments \cite{6}. In the simulation of the molecular propagation channel, the time for detecting the number of arrived information molecules is set as $d/v$, and the duration of a time slot $t_s$ is set as $2d/v$. To avoid gradient explosion during the training process of the channel network, the received symbol vectors are normalized to $[0,1]$ by a min-max linear normalization. 
The performance of the molecular semantic communication system is evaluated using CIFAR-10 image dataset with a classification task. Details are illustrated later.

\begin{table}[!t]
\centering
\caption{Parameters of the Molecular Propagation Channel.}
\label{Molecular_channel_parameters}
\begin{tabular}{ccc}\toprule
Parameters & Scenario 1 & Scenario 2\\\midrule
Radius of the receiver ($r$) & 20 $\mu m$ & 20 $\mu m$\\
Distance ($d$) & 100 $\mu m$ & 20 $m$\\
Diffusion coefficient ($D_c$) & 800 $\mu m^2 /s$ & 800 $\mu m^2 /s$\\
Flow velocity ($v$) & 40 $\mu m/s$ & 10 $m/s$ \\
Time slot ($t_s$) & 5 $s$ & 4$s$
\\\bottomrule
\end{tabular}
\end{table}

\subsection{validatoin of the proposed channel network}
\label{Channel_network}
\begin{figure}[!t]
    \centering
    \includegraphics[height = 8cm]{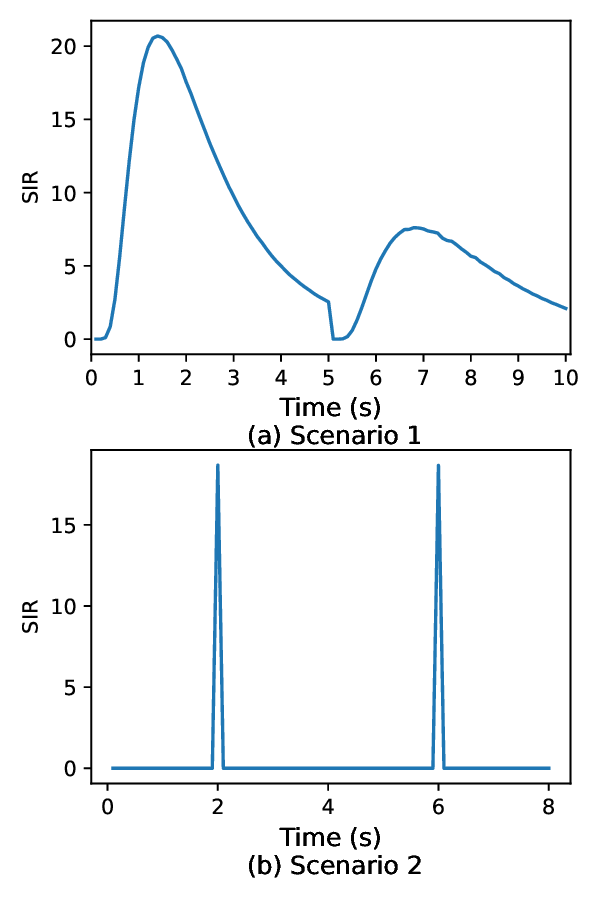}
    \caption{SIR values for sending two consecutive bits $``1"$ in a molecular propagation channel with $n_m$=10000, (a) Scenario 1 (Channel with a low flow velocity) (b) Scenario 2 (Channel with a high flow velocity).}
    \label{SIR_t}
\end{figure}
The temporal variation of SIR during the continuous transmission of two consecutive ``1'' bits through the BCSK modulation system is depicted in Fig.~\ref{SIR_t}. In the simulation, the maximum number of released molecules in each slot is fixed $n_m=10000$. The background Gaussian noise is modeled as a normal distribution with a mean of 1. When the flow velocity is high, as depicted in Fig.~\ref{SIR_t} (b), the propagation of information molecules results in negligible ISI in the channel.
Conversely, for scenarios with low flow velocity, the high ISI will result in a small SIR. As an example, the maximum SIR of the second time slot in Fig.~\ref{SIR_t} (a) is reduced by half because of ISI when compared to the first time slot.

\begin{figure}[!t]
    \centering
    \hspace{-1cm}
    \includegraphics[height = 6cm,width=0.45\textwidth]{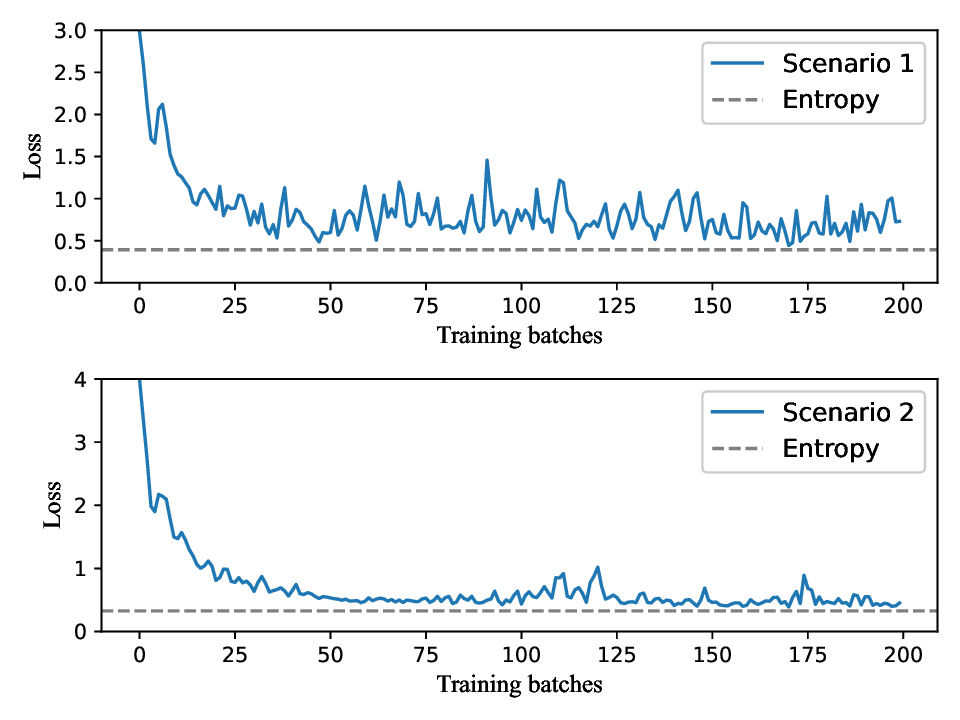}
    \caption{Training loss of the channel network.}
    \label{Channelloss}
\end{figure}

\begin{figure}[!t]
    \centering
    \includegraphics[width=0.5\textwidth]{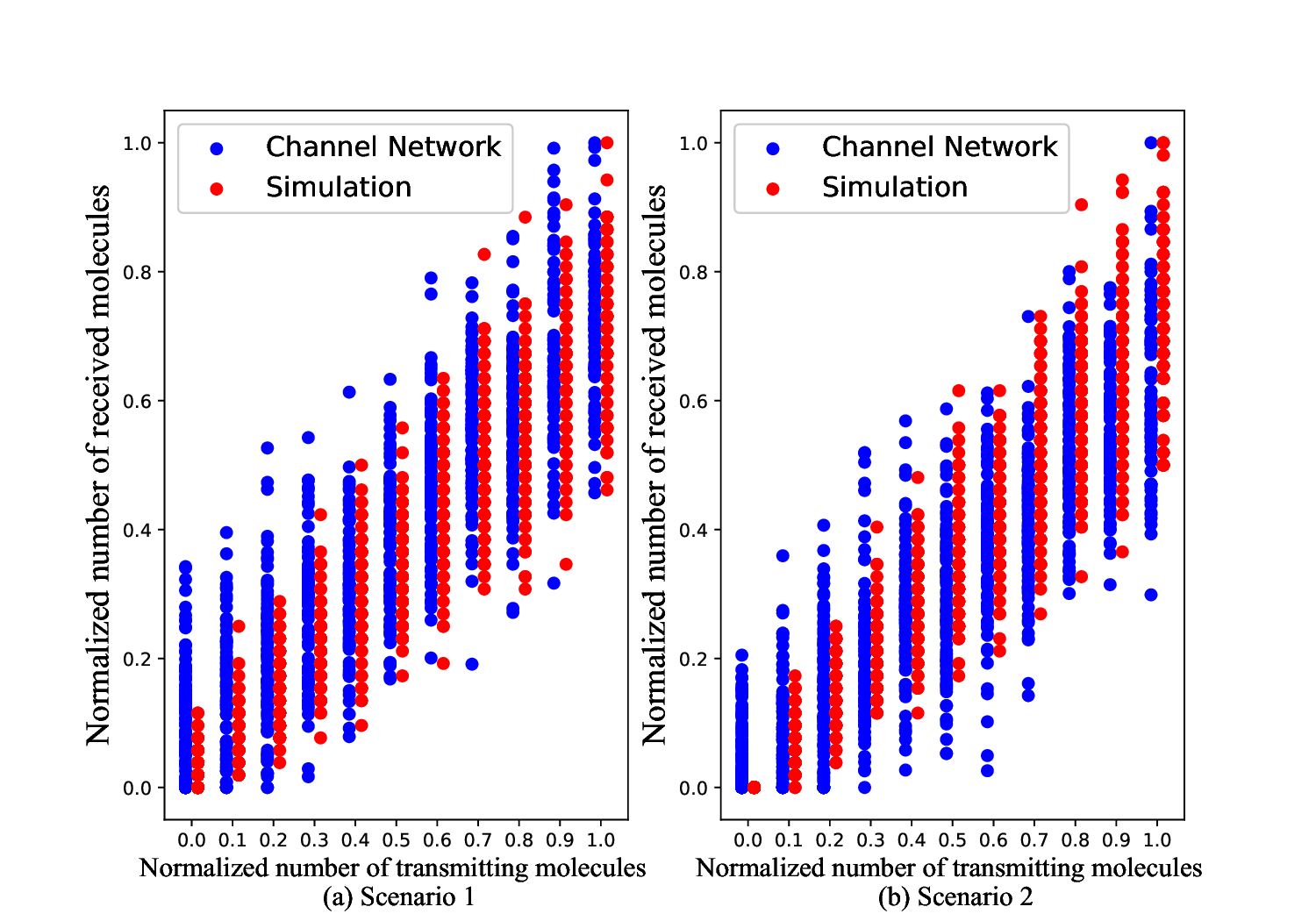}
    \caption{Samples from the channel network and the simulation of molecular propagation channel of two scenarios.}
    \label{Channelsamples}
\end{figure}
We train the channel network with randomly generated symbol vectors of length 100, with a batch size of 20. The training loss of the channel network is evaluated verses the number of vectors.
The convergence of the training loss is shown in Fig.~\ref{Channelloss}. According to \cite{19}, the training loss function in (\ref{Channel_network_loss}) converges to the entropy of the distribution of the normalized received symbol vectors attained from simulation, which is calculated as $H(\hat{\boldsymbol{s}}|\boldsymbol{s}) = -\frac{1}{2} \log(2\pi e \sigma^2),$ where $\sigma^2$ represents the variance of the normalized received vectors.

In Fig.~\ref{Channelsamples}, we illustrate the normalized number of received molecules generated from the channel network and the channel simulation. Each dot represents data sampled from the evaluated distribution under a given normalized number of transmitting molecules.
The channel network demonstrates superior fitting performance when evaluating higher molecular concentrations. However, its effectiveness is limited when the molecular concentrations are lower. The fitting would be better if one uses some analytical function under low molecular concentration, while the specialized function design is beyond the scope of this work.

\subsection{Performance evaluation in classification tasks}
\label{Simulation-results}
In this subsection, we evaluated our proposed method using a classification task. Semantic information of the image is extracted and encoded into molecular concentrations via a JSCC encoder. Upon transmission, the semantic information is decoded from the detected molecular concentrations to accomplish the classification task.
\subsubsection{Experimental setup}
The architecture of the JSCC network and channel network is described in Section \ref{s3}, where the channel network is only used during the training process. The parameters of the molecular propagation channel are shown in Table \ref{Molecular_channel_parameters}. We employ the CIFAR-10 image dataset for training the JSCC network. The dataset consists of 60000 RGB images divided into 10 different classes. Each class includes 6000 images of size 32 × 32 pixels, 5000 for training, and 1000 for testing.
The training data for the channel network are 1,000 randomly generated data as introduced in Section \ref{s33}.
The experimental setup of the system is shown in Table \ref{Exp_setup}.

For comparison, a conventional separate source-channel coding is employed, where joint photographic experts group (JPEG) and low-density parity-check code (LDPC) are adopted as source coding and channel coding, respectively. The adopted modulation scheme is BCSK, where a hard decision of a threshold $n_mP(t_s/2)$ is used to generate the received bits stream. To tackle the ISI, a Minimum Mean Square Error-based equalizer (MMSEE) is adopted. After decoding the JPEG image, a classification network, whose size is similar to that of the semantic JSSC network, is employed for conducting the classification task.

Akin to \cite{15}, the image size $n$, the channel input size $k$, and $R=k/n$ are called the source bandwidth, the channel bandwidth, and the bandwidth compression ratio, respectively.
For the proposed semantic molecular communication method, by adjusting $c_{out}$, which is the number of filters of the last convolutional layer in the encoder, different bandwidth compression ratios can be achieved. In our experiments, networks with different compression ratios $R=2/3, 1/3$ and different maximum number of released molecules $n_m=5000,10000,15000$ are trained and evaluated.
\begin{table}[!t]
\centering
\caption{Data Size for Different Schemes.}

\label{Compression_data_size}
\resizebox{0.45\textwidth}{!}{%
\begin{tabular}{c|c|c|c}\toprule
 & JSCC (R=1/3) & JSCC (R=2/3) & JPEG + 1/2LDPC\\\midrule
Original data size & $32 \times 32 \times 3$ & $32 \times 32 \times 3$ & $32 \times 32 \times 3$ \\
Compressed data size & $16 \times 8 \times 8$ & $32 \times 8 \times 8$ & $\geq 2048$ \\
Channel signal size & $16 \times 8 \times 8$ & $32 \times 8 \times 8$ & $\geq 2048 \times 8$ \\
Bandwidth Compression Ratio & $1/3$ &$ 2/3$ & $\geq 16/3$ 

\\\bottomrule
\end{tabular}%
}
\end{table}
As shown in Table \ref{Compression_data_size}, the data size encoded using JPEG and LDPC slightly exceeds 2/3 of the original image size, which is the maximum bandwidth compression ratio of the JSCC-based schemes. However, due to BCSK modulation, the $R$ of the comparative scheme increases to 16/3, significantly surpassing that of JSCC-based schemes. This concern can be mitigated by employing higher-level modulation schemes \cite{13}. In our experiments, we focus on the BCSK modulation due to its widespread application in other contemporary works (\cite{10130469, 10312770, 9187645}, etc.). The $(288, 576)$ LDPC of rate $1/2$ is employed in the experiments. The simulation is performed by a computing platform with Intel Xeon Silver 4110 CPU 2.10GHz and NVIDIA RTX A4000.

\subsubsection{Simulation results}
\begin{table}[!t]
\centering
\caption{Experimental Setup for Classification Task.}
\label{Exp_setup}
\begin{tabular}{ccc}\toprule
 & JSCC Network & Channel Network\\\midrule
Data & CIFAR-10 & Randomly generated \\
Batch Size & 128 & 50\\
Optimizer & Adam& Adam \\
Learning rate & 0.001& 0.01
\\\bottomrule
\end{tabular}
\end{table}

\begin{figure}[t]
    \centering
    \includegraphics[width=0.45\textwidth]{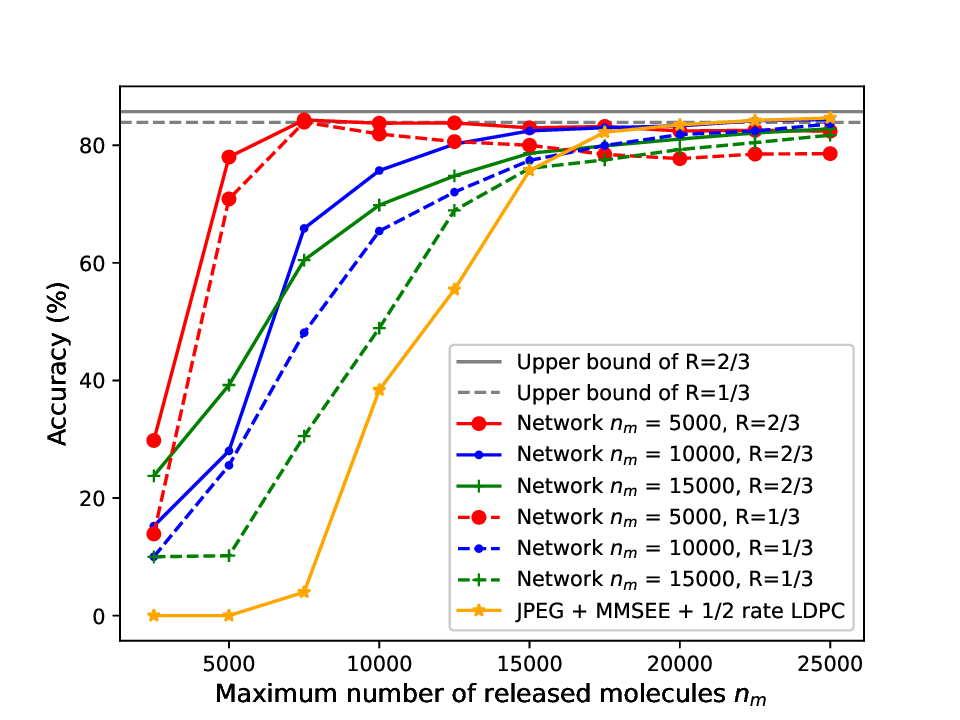}
    \caption{Accuracy performance vs. Maximum molecule number for each slot in Scenario 1 (channel with a low flow velocity).}
    \label{Classification1}
\end{figure}
\begin{figure}[t]
    \centering
    \includegraphics[width=0.45\textwidth]{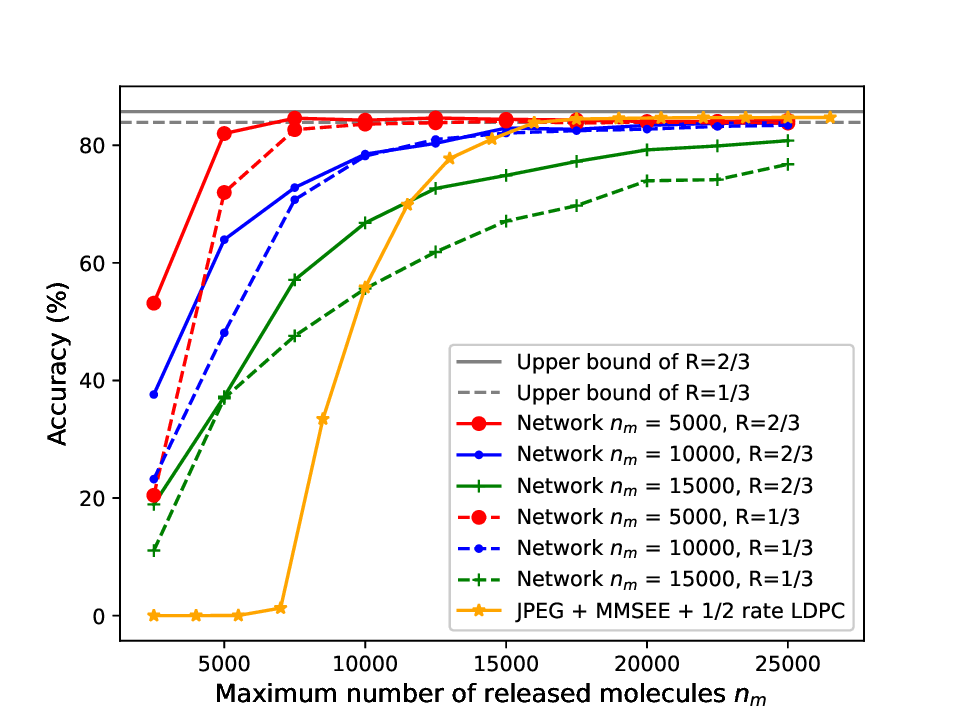}
    \caption{Accuracy performance vs. Maximum molecule number for each slot in Scenario 2 (channel with a high flow velocity).}
    \label{Classification2}
\end{figure}

According to Eq.~(\ref{SIR}), the SIR increases as $n_m$ increases at a certain time $t$. In our experiments, we set the time of detecting arrived molecules to $d/v$, then increasing the $n_m$ is equivalent to increasing the SIR.
The classification performance with a varying $n_m$, i.e., the maximum number of released molecules in each slot for the two scenarios are shown in Fig.~\ref{Classification1} and Fig.~\ref{Classification2}.

The performance of different methods is evaluated under a varying number of maximum released molecules $n_m$. The upper bound curve shows the performance of the proposed network under an ideal channel with no noise and no ISI.
Compared to the conventional method, i.e., JPEG source coding and LDPC channel coding, the proposed semantic JSCC network achieves superior accuracy, particularly at low $n_m$ values. The neural JSCC design helps to avoid the $``$cliff effect$"$ associated with the conventional method, where the serve degradation at low $n_m$ values limits the performance of the conventional method.

Furthermore, we can observe in both scenarios that for the same bandwidth compression ratio $R$, networks trained with lower $n_m$ values in training exhibit better performance. According to \cite{17}, the effect of random Brownian motion in a molecular propagation channel is enhanced for smaller $n_m$. This finding suggests that the parameters of the proposed methods are learned more effectively under high ISI and degradation, resulting in enhanced performance.
Moreover, networks with a higher bandwidth compression ratio exhibit better classification accuracy.

\section{conclusion}
\label{CONCLUSION}

In this paper, we propose a deep JSCC-based semantic molecular communication system, which significantly saves the transmission resource. In this system, the encoder directly maps the task-relevant information to the concentration of information molecules. The encoder and decoder are jointly optimized over the molecular communication channel. To bypass the difficulty brought by the non-differentiable channel, we exploit a channel network in the training stage, which learns an explicit model of the molecular propagation channel.
The proposed JSCC-based semantic molecular communication outperforms the conventional method, i.e., JPEG source coding and LDPC channel coding, especially in the case of a limited number of released molecules per slot, where the system suffers from serve ISI and degradation in the number of detected molecules. It is also found that the parameters of the proposed networks are learned more effectively under high ISI, resulting in improved performance. 


\bibliographystyle{unsrt}  
\bibliography{myref}

\begin{thebibliography}{10}

\bibitem{1}
Claude~Elwood Shannon and Warren Weaver.
\newblock {\em The Mathematical Theory of Communication}.
\newblock University of illinois Press, 1949.

\bibitem{2}
Cheng-Xiang Wang, Xiaohu You, Xiqi Gao, Xiuming Zhu, Zixin Li, Chuan Zhang, Haiming Wang, Yongming Huang, Yunfei Chen, Harald Haas, John~S. Thompson, Erik~G. Larsson, Marco~Di Renzo, Wen Tong, Peiying Zhu, Xuemin Shen, H.~Vincent Poor, and Lajos Hanzo.
\newblock On the road to 6g: Visions, requirements, key technologies, and testbeds.
\newblock {\em IEEE Communications Surveys $\&$ Tutorials}, 25(2):905--974, 2023.

\bibitem{3}
Deniz Gündüz, Zhijin Qin, Inaki~Estella Aguerri, Harpreet~S. Dhillon, Zhaohui Yang, Aylin Yener, Kai~Kit Wong, and Chan-Byoung Chae.
\newblock Beyond transmitting bits: Context, semantics, and task-oriented communications.
\newblock {\em IEEE Journal on Selected Areas in Communications}, 41(1):5--41, 2023.

\bibitem{10480348}
Tong Wu, Zhiyong Chen, Dazhi He, Liang Qian, Yin Xu, Meixia Tao, and Wenjun Zhang.
\newblock Cddm: Channel denoising diffusion models for wireless semantic communications.
\newblock {\em IEEE Transactions on Wireless Communications}, pages 1--1, 2024.

\bibitem{10107616}
Xinlai Luo, Zhiyong Chen, Meixia Tao, and Feng Yang.
\newblock Encrypted semantic communication using adversarial training for privacy preserving.
\newblock {\em IEEE Communications Letters}, 27(6):1486--1490, 2023.

\bibitem{4}
Tadashi Nakano.
\newblock {\em Molecular Communication}.
\newblock Cambridge University Press, 2013.

\bibitem{5}
Nariman Farsad, H.~Birkan Yilmaz, Andrew Eckford, Chan-Byoung Chae, and Weisi Guo.
\newblock A comprehensive survey of recent advancements in molecular communication.
\newblock {\em IEEE Communications Surveys $\&$ Tutorials}, 18(3):1887--1919, 2016.

\bibitem{6}
Sebastian Lotter, Lukas Brand, Vahid Jamali, Maximilian Schäfer, Helene~M. Loos, Harald Unterweger, Sandra Greiner, Jens Kirchner, Christoph Alexiou, Dietmar Drummer, Georg Fischer, Andrea Buettner, and Robert Schober.
\newblock Experimental research in synthetic molecular communications – part ii.
\newblock {\em IEEE Nanotechnology Magazine}, 17(3):54--65, 2023.

\bibitem{7}
Song Qiu, Weisi Guo, Siyi Wang, Nariman Farsad, and Andrew Eckford.
\newblock A molecular communication link for monitoring in confined environments.
\newblock In {\em 2014 IEEE International Conference on Communications Workshops (ICC)}, pages 718--723, 2014.

\bibitem{8}
Weisi Guo, Christos Mias, Nariman Farsad, and Jiang-Lun Wu.
\newblock Molecular versus electromagnetic wave propagation loss in macro-scale environments.
\newblock {\em IEEE Transactions on Molecular, Biological and Multi-Scale Communications}, 1(1):18--25, 2015.

\bibitem{9}
Weisi Guo, Iresha Atthanayake, and Peter Thomas.
\newblock Vertical underwater molecular communications via buoyancy: Gaussian velocity distribution of signal.
\newblock In {\em ICC 2020 - 2020 IEEE International Conference on Communications (ICC)}, pages 1--6, 2020.

\bibitem{10}
L.~Felicetti, M.~Femminella, G.~Reali, and P.~Liò.
\newblock Applications of molecular communications to medicine: A survey.
\newblock {\em Nano Communication Networks}, 7:27--45, 2016.

\bibitem{11}
Abdulaziz Al-Helali, Ben Liang, and Nidal Nasser.
\newblock Novel molecular signaling method and system for molecular communication in human body.
\newblock {\em IEEE Access}, 8:119361--119375, 2020.

\bibitem{12}
Samudra Sengupta, Michael~E. Ibele, and Ayusman Sen.
\newblock Fantastic voyage: Designing self-powered nanorobots.
\newblock {\em Angewandte Chemie International Edition}, 51(34):8434--8445, 2012.

\bibitem{13}
Mehmet~$\d{S}\"{u}kr\"{u}$ Kuran, H.~Birkan Yilmaz, Ilker Demirkol, Nariman Farsad, and Andrea Goldsmith.
\newblock A survey on modulation techniques in molecular communication via diffusion.
\newblock {\em IEEE Communications Surveys $\&$ Tutorials}, 23(1):7--28, 2021.

\bibitem{14}
Pit Hofmann, Juan~A. Cabrera, Riccardo Bassoli, Martin Reisslein, and Frank H.~P. Fitzek.
\newblock Coding in diffusion-based molecular nanonetworks: A comprehensive survey.
\newblock {\em IEEE Access}, 11:16411--16465, 2023.

\bibitem{9438648}
Jialong Xu, Bo~Ai, Wei Chen, Ang Yang, Peng Sun, and Miguel Rodrigues.
\newblock Wireless image transmission using deep source channel coding with attention modules.
\newblock {\em IEEE Transactions on Circuits and Systems for Video Technology}, 32(4):2315--2328, 2022.

\bibitem{10225310}
Jialong Xu, Bo~Ai, Wei Chen, Ning Wang, and Miguel Rodrigues.
\newblock Deep joint source-channel coding for image transmission with visual protection.
\newblock {\em IEEE Transactions on Cognitive Communications and Networking}, 9(6):1399--1411, 2023.

\bibitem{9954153}
Jialong Xu, Bo~Ai, Ning Wang, and Wei Chen.
\newblock Deep joint source-channel coding for csi feedback: An end-to-end approach.
\newblock {\em IEEE Journal on Selected Areas in Communications}, 41(1):260--273, 2023.

\bibitem{15}
Eirina Bourtsoulatze, David Burth~Kurka, and Deniz Gündüz.
\newblock Deep joint source-channel coding for wireless image transmission.
\newblock {\em IEEE Transactions on Cognitive Communications and Networking}, 5(3):567--579, 2019.

\bibitem{16}
Jialong Xu, Tze-Yang Tung, Bo~Ai, Wei Chen, Yuxuan Sun, and Deniz Gündüz.
\newblock Deep joint source-channel coding for semantic communications.
\newblock {\em IEEE Communications Magazine}, 61(11):42--48, 2023.

\bibitem{20}
Fayçal~Ait Aoudia and Jakob Hoydis.
\newblock Model-free training of end-to-end communication systems.
\newblock {\em IEEE Journal on Selected Areas in Communications}, 37(11):2503--2516, 2019.

\bibitem{17}
Timothy~J. O’Shea, Tamoghna Roy, and Nathan West.
\newblock Approximating the void: Learning stochastic channel models from observation with variational generative adversarial networks.
\newblock In {\em 2019 International Conference on Computing, Networking and Communications (ICNC)}, pages 681--686, 2019.

\bibitem{18}
Jiarui Zhu, Chenyao Bai, Yunlong Zhu, Xiwen Lu, and Kezhi Wang.
\newblock Evolutionary generative adversarial network based end-to-end learning for mimo molecular communication with drift system.
\newblock {\em Nano Communication Networks}, 37:100456, 2023.

\bibitem{19}
Dolores García, Jesus~O. Lacruz, Damiano Badini, Danilo {De Donno}, and Joerg Widmer.
\newblock Model-free machine learning of wireless siso/mimo communications.
\newblock {\em Computer Communications}, 181:192--202, 2022.

\bibitem{21}
Vahid Jamali, Arman Ahmadzadeh, Wayan Wicke, Adam Noel, and Robert Schober.
\newblock Channel modeling for diffusive molecular communication—a tutorial review.
\newblock {\em Proceedings of the IEEE}, 107(7):1256--1301, 2019.

\bibitem{citeone}
Nariman Farsad, David Pan, and Andrea Goldsmith.
\newblock {A} novel experimental platform for in-vessel multi-chemical molecular communications.
\newblock In {\em GLOBECOM 2017 - 2017 IEEE Global Communications Conference}, pages 1--6, 2017.

\bibitem{citetwo}
Wayan Wicke, Harald Unterweger, Jens Kirchner, Lukas Brand, Arman Ahmadzadeh, Doaa Ahmed, Vahid Jamali, Christoph Alexiou, Georg Fischer, and Robert Schober.
\newblock Experimental system for molecular communication in pipe flow with magnetic nanoparticles.
\newblock {\em IEEE Transactions on Molecular, Biological and Multi-Scale Communications}, 8(2):56--71, 2022.

\bibitem{9027995}
Ladan Khaloopour, Seyed~Vahid Rouzegar, Alireza Azizi, Amir Hosseinian, Maryam Farahnak-Ghazani, Nahal Bagheri, Mahtab Mirmohseni, Hamidreza Arjmandi, Reza Mosayebi, and Masoumeh Nasiri-Kenari.
\newblock {A}n experimental platform for macro-scale fluidic medium molecular communication.
\newblock {\em IEEE Transactions on Molecular, Biological and Multi-Scale Communications}, 5(3):163--175, 2019.

\bibitem{22}
Christopher~M. Bishop.
\newblock Mixture density networks.
\newblock 1994.

\bibitem{24}
Lei Kong, Li~Huang, Lin Lin, Zhimin Zheng, Yu~Li, Qixing Wang, and Guangyi Liu.
\newblock A survey for possible technologies of micro/nanomachines used for molecular communication within 6g application scenarios.
\newblock {\em IEEE Internet of Things Journal}, 10(13):11240--11263, 2023.

\bibitem{10130469}
Valerio Selis, Daniel~Tunç McGuiness, and Alan Marshall.
\newblock A novel ml-based symbol detection pipeline for molecular communication.
\newblock {\em IEEE Transactions on Molecular, Biological and Multi-Scale Communications}, 9(2):207--216, 2023.

\bibitem{10312770}
Mustafa~Can Gursoy and Urbashi Mitra.
\newblock Scheduling-based transmit signal shaping in energy-constrained molecular communications.
\newblock {\em IEEE Transactions on Molecular, Biological and Multi-Scale Communications}, 9(4):447--460, 2023.

\bibitem{9187645}
Trang~Ngoc Cao, Vahid Jamali, Nikola Zlatanov, Phee~Lep Yeoh, Jamie Evans, and Robert Schober.
\newblock Fractionally spaced equalization and decision feedback sequence detection for diffusive mc.
\newblock {\em IEEE Communications Letters}, 25(1):117--121, 2021.

\end{thebibliography}

\end{document}